\documentstyle[preprint,aps]{revtex}

\begin{document}

\draft

\tighten

\preprint{\vbox{\hfill UCSB--TH--94--40 \\
          \vbox{\hfill cond-mat/9410046} \\
          \vbox{\hfill October 1994} \\
          \vbox{\hfill Revised April 1995}
          \vbox{\vskip1.0in}
         }}

\title{Does quantum chaos explain quantum statistical mechanics?}

\author{Mark Srednicki\footnote{E--mail: \tt mark@tpau.physics.ucsb.edu}}

\address{Department of Physics, University of California,
         Santa Barbara, CA 93106
         \\ \vskip0.5in}

\maketitle

\begin{abstract}
\normalsize{
If a many-body quantum system approaches thermal equilibrium from a generic
initial state, then the expectation value $\langle\psi(t)|A_i|\psi(t)\rangle$,
where $|\psi(t)\rangle$ is the system's state vector and $A_i$ is an
experimentally accessible observable, should approach a constant value
which is independent of the initial state, and equal to a thermal average
of $A_i$ at an appropriate temperature.  We show that this is the case
for all simple observables whenever the system is classically chaotic.}
\end{abstract}


Many physicists would agree with Ma's opinion that ``statistical
mechanics, as it is at present, is an ill-proportioned subject, with
many successful applications but relatively little understanding of
the basic principles'' \cite{ma}.  This statement can be disputed for
classical statistical mechanics, since modern chaos theory has
illuminated Boltzmann's ``ergodic hypothesis'' \cite{boltz}
for closed classical systems (see, e.g., \cite{rasetti}).
However, there has been comparatively little work on the foundations
of statistical mechanics for closed quantum systems.
Closed quantum systems are the subject of this Letter;
I demonstrate that results from quantum chaology (the study of
%%%%% Editor please note:  the word "chaology" has appeared previously in PRL.
quantum systems which are classically chaotic) can provide a
sound, easily understood basis for quantum statistical mechanics.

A closed, bounded quantum system has a discrete energy spectrum, and
the state vector of such a system at time $t$ can be written as
\begin{equation}
|\psi(t)\rangle = \sum_\alpha C_\alpha\,e^{-iE_\alpha t/\hbar}\,
                  |\alpha\rangle \, ,                              \label{1}
\end{equation}
where $|\alpha\rangle$ is an energy eigenstate with eigenvalue
$E_\alpha$, and the coefficients $C_\alpha$ (normalized so that
$\sum_\alpha |C_\alpha|^2=1$) specify the initial state.  We would
like to understand how an approach to thermal equilibrium can be
encoded in Eq.\ (\ref{1}).

To decide whether a system is in thermal equilibrium, we must measure
something.  Let us say that the quantities to be measured are
represented by a set of hermitian operators $A_i$, and that our
measurements can be characterized by some finite time average of the
quantum expectation values $\langle\psi(t)|A_i|\psi(t)\rangle$.  This
is clearly a drastic oversimplification of any real experiment, but we
will make the assumption that a detailed analysis of the measurement
process is not essential to an understanding of statistical mechanics.

Although we cannot measure them, an interesting set of objects to
consider are the {\em infinite\/} time averages,
\begin{eqnarray}
\overline{A_i} &=& \lim_{\tau\to\infty}{1\over\tau}
                   \int_0^\tau dt\;\langle\psi(t)|A_i|\psi(t)\rangle
                                                         \nonumber \\
\noalign{\bigskip}
               &=& \sum_\alpha |C_\alpha|^2\,
                   \langle\alpha|A_i|\alpha\rangle \, .           \label{2}
\end{eqnarray}
If our  system approaches thermal equilibrium from a generic initial state,
then we would expect $\overline{A_i}$ to depend on the total energy of the
system, but to be independent of all other aspects of the initial
state \cite{peres}.  This is because, even though we will have some nonthermal
behavior at the beginning (which will of course depend on the initial state),
after the system reaches thermal equilibrium, it should stay there except
for small (or large but rare) fluctuations.  All of these short-time effects
should be washed out by the infinite time average.  Even so, the system will
have a well defined final temperature only if its total energy
$\langle E\rangle =\sum_\alpha |C_\alpha|^2E_\alpha$ has a quantum uncertainty
$\Delta E = (\sum_\alpha |C_\alpha|^2 E_\alpha^2 - \langle E\rangle^2){}^{1/2}$
which is small:  $\Delta E \ll \langle E\rangle$.  From now on we will restrict
our attention to initial states which satisfy this condition.  Returning to
Eq.\ (\ref{2}), we can see that $\overline{A_i}$ will be independent of the
detailed pattern of the $C_\alpha$'s if and only if
$\langle\alpha|A_i|\alpha\rangle$ is a smooth function of $E_\alpha$,
with negligible variation over the relatively small range
$\langle E\rangle\pm\Delta E$.  Whether or not
this is a reasonable condition on the $A_i$'s will be taken up later.

For now, let us assume $\langle\alpha|A_i|\alpha\rangle$ is indeed a
smooth function of $E_\alpha$, and write it as
$\langle\alpha|A_i|\alpha\rangle = \Phi_i(E_\alpha)$.
Thus we have
\begin{equation}
\overline{A_i} = \Phi_i(\langle E\rangle)\,
                 [1 + O(\Delta E/\langle E\rangle)] \, .           \label{3}
\end{equation}
We can gain some understanding of the function $\Phi_i(E)$ if we
consider the {\em thermal\/} expectation value
\begin{eqnarray}
\langle A_i\rangle_T &=& Z(T)^{-1} \sum_\alpha e^{-E_\alpha/T}
                         \langle\alpha|A_i|\alpha\rangle \nonumber \\
\noalign{\bigskip}
                     &=& Z(T)^{-1} \int_0^\infty dE\
                         n(E)\,e^{-E/T}\,\Phi_i(E) \, ,            \label{4}
\end{eqnarray}
where $T$ is the temperature (and we have set Boltzmann's constant to one),
$Z(T)=\sum_\alpha e^{-E_\alpha/T}$ is the partition function,
and $n(E) = \sum_\alpha \delta(E-E_\alpha)$ is the density of states.
For a system with $N\gg 1$ degrees of freedom,
and when integrated against a smooth function of $E$,
the exact $n(E)$ can be replaced (with negligible error) by a smooth
function of the form $\bar n(E) \sim \exp[Ns(E/N)]$,
where $s(E/N)$ can be identified as the thermodynamic entropy per
degree of freedom.  We then have
\begin{eqnarray}
\langle A_i\rangle_T &=&
       {\int_0^\infty dE\ e^{N[s(E/N)-E/NT]}\,\Phi_i(E)
        \over
        \int_0^\infty dE\ e^{N[s(E/N)-E/NT]} } \nonumber \\
\noalign{\bigskip}
             &=& \Phi_i(U)\,[1 + O(N^{-1/2})] \, ,                  \label{5}
\end{eqnarray}
where the second line follows from the steepest-descent approximation,
and $U(T)=T^2 Z'(T)/Z(T)$ is the thermodynamic internal energy, given
by the solution of the steepest-descent condition $s'(U/N) = 1/T$.
Comparing Eqs.\ (\ref{3}) and (\ref{5}), and continuing to assume both
$\Delta E \ll \langle E\rangle$ and $N \gg 1 $, we obtain a satisfying
result:  the infinite time average $\overline{A_i}$ is equal to the
thermal average $\langle A_i\rangle_T$, with the temperature $T$ related
to the quantum expectation value of the energy $\langle E\rangle$ by
$U(T)=\langle E\rangle$.

Furthermore, since $\langle\alpha|A_i|\alpha\rangle=\Phi_i(E_\alpha)$,
Eq.\ (\ref{5}) implies that
$\langle\alpha|A_i|\alpha\rangle=\langle A_i\rangle_{T_\alpha}$,
where $T_\alpha$ is given by the solution of $U(T_\alpha)=E_\alpha$.
Thus, if $\overline{A_i}$ is to be independent of the initial state,
then individual energy eigenstates must have thermal properties.
This phenomenon was previously discovered (via different arguments)
in the special case of the hard-sphere gas, where it was dubbed
{\em eigenstate thermalization} \cite{me}.

Now let us come back to consider
\begin{equation}
\langle\psi(t)|A_i|\psi(t)\rangle
    = \sum_{\alpha\beta}C^*_\alpha C^{\phantom{*}}_\beta\,
    e^{i(E_\alpha-E_\beta)t/\hbar}\langle\alpha|A_i|\beta\rangle \label{6}
\end{equation}
without any time averaging at all.  We expect that
$\langle\psi(t)|A_i|\psi(t)\rangle$ will start out at $t=0$ with
whatever value is picked out by the initial state, and then evolve to
its thermal expectation value $\langle A_i\rangle_T$ after some
characteristic relaxation time.  Examining Eq.\ (\ref{6}), we see that
this can only happen if $\langle\alpha|A_i|\beta\rangle$ is
generically much smaller than $\langle\alpha|A_i|\alpha\rangle$ when
$\beta\ne\alpha$.  In this case, the $\beta\ne\alpha$ terms in
Eq.\ (\ref{6}) will usually make a negligible contribution, and then we
will find $\langle\psi(t)|A_i|\psi(t)\rangle=\langle A_i\rangle_T$,
just as we found $\overline{A_i}=\langle A_i\rangle_T$.  However, if
the magnitudes and phases of the $C_\alpha$'s are carefully chosen,
then we can ``line up'' the $\langle\alpha|A_i|\beta\rangle$'s so as
to get any value of $\langle\psi(0)|A_i|\psi(0)\rangle$ that we might
want.  Afterward, however, as we see in Eq.\ (\ref{6}), the phases will
change in the usual manner; the carefully contrived coherence among
the various $\langle\alpha|A_i|\beta\rangle$'s will eventually be destroyed,
and we will again find
$\langle\psi(t)|A_i|\psi(t)\rangle=\langle A_i\rangle_T$.
The outstanding problem, not to be solved here, is to understand this
process more quantitatively for specific systems.
This will entail characterizing classes of physically relevant initial states
by some sort of statistical relationship between the $C_\alpha$'s and
the $\langle\alpha|A_i|\beta\rangle$'s, and then demonstrating that
this relationship results in the appropriate phenomenological equations
(e.g., the Boltzmann equation for a dilute gas).  This program is obviously
an ambitious one.

We can, however, use our present results to resolve a paradox \cite{mermin}
which arises in the conventional approach to quantum statistical
mechanics.  Suppose we have a Carnot engine \cite{carnot}:
a system with an adjustable parameter $V$ (such as its volume,
or an external electric or magnetic field) which has been brought
to thermal equilibrium by placing it in contact with a large heat bath
at temperature $T_1$.  This means that we should describe
the system with a ``thermal'' density matrix that is diagonal in the
energy eigenstate basis, with diagonal entries given by the Boltzmann
weights $Z(T_1,V_1)^{-1}e^{-E_\alpha(V_1)/T_1}$, where $V_1$ is the
value of the parameter $V$.  Now remove the system from contact with
the bath, and change $V$ very slowly from $V_1$ to $V_2$.  According
to thermodynamics, this is an ``adiabatic'' process, and the system
remains in thermal equilibrium at a slowly varying temperature.
At the end of the process, conventional statistical mechanics would
assign the system a new thermal density matrix, diagonal in the
energy eigenstate basis with diagonal entries
$Z(T_2,V_2)^{-1}e^{-E_\alpha(V_2)/T_2}$, where $T_2$ is the final
temperature.  However, according to quantum mechanics, a sufficiently
slow (or ``adiabatic'') process preserves the probability for the
system to be in any given energy eigenstate (although the eigenstates
themselves change with $V$); thus the system does {\em not\/} acquire
the new Boltzmann weights appropriate to its new volume and
temperature, but instead retains the original ones.  How, then,
can we understand it to be in thermal equilibrium?

The conventional response is to declare the question irrelevant,
either because in practice we cannot change $V$ slowly enough
to invoke quantum mechanical adiabaticity \cite{ma}, or because
we can only ``essentially'' but not ``completely''
isolate a system from its environment \cite{tolman}.

Our answer is different, and has already been presented:
a system which exhibits eigenstate thermalization can come to thermal
equilibrium for {\em any\/} initial state with a relatively small
energy uncertainty.
The discussion was for pure states, but it is trivially generalized
to include mixed states:
simply replace $C^*_\alpha C^{\phantom{*}}_\beta$ by $\rho_{\alpha\beta}$
in Eq.\ (\ref{6}), where $\rho_{\alpha\beta}$ is the initial density matrix
in the energy eigenstate basis.
It follows immediately from our previous considerations
that if $\rho_{\alpha\beta}$ is diagonal,
and if $\Delta E/\langle E\rangle$ is small,
then the system is automatically in thermal equilibrium.
In the case at hand, the initial thermal density matrix (which
according to quantum mechanics is unchanged during the adiabatic
process) certainly meets these two conditions:  it is explicitly
diagonal, and for a system with $N\gg 1$ degrees of freedom we know
(from conventional statistical mechanics)
that $\Delta E \sim N^{-1/2}\langle E\rangle$.
However, the energy eigenstates {\em do\/} change with $V$, and so
their thermal properties reflect the new volume $V_2$.  Meanwhile,
the expectation value of the energy changes because the energy
eigenvalues do, and this results in a new temperature.  All together,
this means that, after the slow change in $V$ has been completed,
the quantum expectation value of each $A_i$ will be equal to the thermal
expectation value that is predicted (for volume $V_2$ and temperature $T_2$)
by conventional statistical mechanics, even though each energy eigenstate
retains its original Boltzmann weight.
In short, the paradox is resolved by eigenstate thermalization.

All of our results depend crucially on two assumptions: that
$\langle\alpha|A_i|\alpha\rangle$ is a smooth function of $E_\alpha$, and that
$|\langle\alpha|A_i|\beta\rangle| \ll |\langle\alpha|A_i|\alpha\rangle|$ when
$\beta\ne\alpha$.  Under what circumstances can we expect these assumptions
to hold? A comprehensive answer is of course out of reach, since the only
restrictions we have placed on the systems under consideration are that they
are bounded (so that the energy spectrum is discrete) and that they have
many degrees of freedom.  Nevertheless, it is possible to identify a broad
class of systems for which the validity of our assumptions can be
demonstrated, in some cases with full mathematical rigor:  many-body
quantum systems which are classically chaotic.

For such a system, one can make a semiclassical calculation of the matrix
elements (in the energy eigenstate basis) of any operator $A_i(p,q)$ which
does not depend explicitly on $\hbar$, and is a smooth function of the
canonical coordinates and momenta.  The diagonal matrix elements
are given by \cite{shnir,nord,voros,b77a,b77b,zeld,cdv}
\begin{equation}
\langle\alpha|A_i|\alpha\rangle
       = \{A_i\}_{E_\alpha}\,[1+O(\hbar)] \, ,                    \label{7}
\end{equation}
where we have defined the classical, microcanonical average
\begin{equation}
\{A_i\}_E = {\int d^N\!p\,d^N\!q\,\delta(E-H(p,q))A_i(p,q)
                              \over
             \int d^N\!p\,d^N\!q\,\delta(E-H(p,q))} \, ,          \label{8}
\end{equation}
and $H(p,q)$ is the classical hamiltonian.  Eq.\ (\ref{7}) has been
proven rigorously when the chaotic system consists of geodesic motion of
a particle on a compact $N$-dimensional manifold with everywhere negative
curvature \cite{shnir,zeld,cdv}; the $O(\hbar)$ term vanishes as
$E_\alpha \to \infty$.  Clearly this means that
$\langle\alpha|A_i|\alpha\rangle$ is a smooth function of $E_\alpha$ for
sufficiently large values of $E_\alpha$.  We can, however, go further; we will
see that, for any particular eigenstate $|\alpha\rangle$, the expected
deviation $\Delta\langle\alpha|A_i|\alpha\rangle$ of
$\langle\alpha|A_i|\alpha\rangle$ from the estimate of Eq.\ (\ref{7}) is
exponentially small in $N$, and hence numerically negligible when $N\gg 1$.
First, though, we turn our attention to the off-diagonal matrix elements
$\langle\alpha|A_i|\beta\rangle$.

We begin with the observation that, for a quantized chaotic system, the
numerical values of the off-diagonal matrix elements
$\langle\alpha|A_i|\beta\rangle$ should be smoothly distributed when
$E_\alpha$ and $E_\beta$ are in fixed, narrow, nonoverlapping ranges.
The idea here is that, since there are no classically conserved or
approximately conserved quantities other than the total energy,
there should be no selection rules which would make some matrix elements
much larger than others \cite{perc,pech,peres2}.  A smooth distribution
for the $\langle\alpha|A_i|\beta\rangle$'s also follows from Berry's
conjecture that the energy eigenfunctions behave like gaussian random
variables \cite{b77b,blec,stein,lirob}.  Next we note that if
$A_i$ is to be a useful observable, its quantum uncertainty $\Delta A_i$
cannot be too large.  For an energy eigenstate $|\alpha\rangle$, $\Delta A_i$
is given by \cite{pech}
\begin{eqnarray}
(\Delta A_i)^2 &=& \langle\alpha|A_i^2|\alpha\rangle -
                   \langle\alpha|A_i|\alpha\rangle^2     \nonumber \\
\noalign{\bigskip}
               &=& \sum_{\beta\ne\alpha}|\langle\alpha|A_i|\beta\rangle|^2
                                                         \nonumber \\
\noalign{\smallskip}
               &\sim& {\bar n}(E_\alpha)\,E_\alpha\,
                      |\langle\alpha|A_i|\beta\rangle|^2 \, .       \label{9}
\end{eqnarray}
In the second line, we used completeness, and in the third, crudely
approximated the sum; in the third line, $|\langle\alpha|A_i|\beta\rangle|^2$
stands for a typical value of the squared off-diagonal matrix element when
$E_\beta$ is close to $E_\alpha$.  The key point is that ${\bar n}(E)$
is exponentially large in $N$, while $\Delta A_i$ cannot be very much larger
than $\langle\alpha|A_i|\alpha\rangle$.  Thus, the typical value of
$|\langle\alpha|A_i|\beta\rangle|^2$ must be exponentially small in $N$.

This argument can be considerably refined \cite{fp,proz};
a semiclassical estimate of the typical value of
$|\langle\alpha|A_i|\beta\rangle|^2$ is
\begin{equation}
|\langle\alpha|A_i|\beta\rangle|^2
    = {S_i(\omega,E)\over \hbar{\bar n}(E)}\,[1+O(\hbar)] \, ,     \label{10}
\end{equation}
where $\omega=(E_\alpha-E_\beta)/\hbar$ and $E={1\over2}(E_\alpha+E_\beta)$,
and we have defined the power spectrum $S_i(\omega,E)$ of the classical
observable $A_i$ on the phase-space surface of energy $E$,
\begin{equation}
S_i(\omega,E) = {1\over 2\pi}\int_{-\infty}^{+\infty}dt\,e^{i\omega t}\,
                \bigl(\{A_i(t)A_i(0)\}_E - \{A_i(0)\}^2_E \bigr)\,.\label{11}
\end{equation}
Again, the key point is that $S_i(\omega,E)$ can reflect
only the intrinsic $N$ dependence of $A_i$ (e.g., extensive or intensive),
while ${\bar n}(E)$ grows exponentially with $N$; thus the
off-diagonal matrix elements must be exponentially small in $N$.
This conclusion must hold even after all quantum corrections to
Eq.\ (\ref{10}) are included.  (Also note that the factor of $1/\hbar$
on the right-hand side of Eq.\ (\ref{10}) does not mean that the off-diagonal
matrix elements become large when $\hbar$ is small, since $1/{\bar n}(E)$ is
proportional to $\hbar^N$.)

Now we return to our discussion of the diagonal matrix elements
$\langle\alpha|A_i|\alpha\rangle$.  For any particular eigenstate
$|\alpha\rangle$, the expected deviation
$\Delta\langle\alpha|A_i|\alpha\rangle$ of $\langle\alpha|A_i|\alpha\rangle$
from the estimate of Eq.\ (\ref{7}) is given by
the $\omega\to 0$ limit of Eq.\ (\ref{10}) \cite{fp}; that is,
\begin{equation}
[\Delta\langle\alpha|A_i|\alpha\rangle]^2
    = {S_i(0,E_\alpha)\over \hbar{\bar n}(E_\alpha)}\,[1+O(\hbar)]\,.\label{12}
\end{equation}
The previous discussion of $N$ dependence applies, and so we see that
$\Delta\langle\alpha|A_i|\alpha\rangle$ is exponentially small in $N$.
Once again, this conclusion must hold even when all quantum corrections
to Eqs.\ (\ref{7}) and (\ref{12}) are included.  Thus, for a
quantized chaotic system with many degrees of freedom, we expect
$\langle\alpha|A_i|\alpha\rangle$ to be a smooth function of $E_\alpha$
for any operator $A_i(p,q)$ which is a smooth function of $p$ and $q$,
and we also expect
$|\langle\alpha|A_i|\beta\rangle| \ll |\langle\alpha|A_i|\alpha\rangle|$
when $\beta\ne\alpha$.  These are precisely the conditions we need in order
to understand the emergence of statistical mechanics from the underlying
quantum dynamics.

To conclude, we have seen that a closed quantum system with many degrees
of freedom can approach thermal equilibrium from a generic (but definite)
initial state if the observables $A_i$ which are measured to check for
thermal equilibrium satisfy two conditions.  First, the expectation value
$\langle\alpha|A_i|\alpha\rangle$, where $|\alpha\rangle$ is an energy
eigenstate, must be a smooth function of the energy eigenvalue $E_\alpha$.
Second, the off-diagonal matrix elements $\langle\alpha|A_i|\beta\rangle$
must be much smaller than the corresponding diagonal ones.  If the first
condition is met, then $\langle\alpha|A_i|\alpha\rangle$ turns out to be equal
to the {\em thermal\/} expectation value of $A_i$ at a temperature $T_\alpha$
given by $U(T_\alpha)=E_\alpha$, where $U(T)$ is the thermodynamic internal
energy of the system (computed in the usual way) as a function of temperature;
this is {\em eigenstate thermalization}.  If the second condition is also
met, then we have a basis for understanding how a system which is prepared
in a nonequilibrium state can evolve to thermal equilibrium via hamiltonian
time evolution.  Both conditions are expected to be satisfied by all simple
observables whenever the quantum system is obtained by quantization of a
many-body classical system which is chaotic.
Of course, the two conditions could also be satisfied
even if the quantum hamiltonian does {\em not\/} have a classical limit
(e.g., a lattice spin system, perhaps with the requirement that the
$A_i$'s are local or bilocal).  An interesting open question is whether
or not quantum systems of this type have any other properties in common
(such as eigenvalue spacing statistics \cite{space})
which would serve to identify them more easily.

\begin{acknowledgments}

I would like to thank David Mermin for telling me about the paradox
which I claim to solve, and Walter Kohn for bringing the work of
Asher Peres to my attention.  This work was supported in part by NSF
Grant PHY--91--16964.

\end{acknowledgments}

\end{document}